\documentclass{iau}
\pdfoutput=1
\usepackage{graphicx}

\newcommand{\void}{\mathrm{v}}

\newcommand{\rs}{r_s}
\newcommand{\dc}{\delta_c}

\title[Modeling cosmic void statistics] 
{Modeling cosmic void statistics}

\author[Nico Hamaus, P.~M. Sutter \& Benjamin~D. Wandelt]   
{Nico Hamaus$^{1,2,\thanks{email: {\tt hamaus@iap.fr}}}$,
P.~M. Sutter$^{1,2,3}$, and Benjamin~D. Wandelt$^{1,2,4}$}

\affiliation{$^1$Sorbonne Universit\'es, UPMC Univ Paris 06, Paris, France \\[\affilskip]
$^2$CNRS, UMR 7095, Institut d'Astrophysique de Paris, Paris, France \\[\affilskip]
$^3$Center for Cosmology and AstroParticle Physics, Ohio State University, Columbus, USA \\[\affilskip]
$^4$Departments of Physics and Astronomy, University of Illinois at Urbana-Champaign, USA
}

\pubyear{2014}
\volume{308}  
\pagerange{0--0}
\jname{The Zeldovich Universe: Genesis and Growth of the Cosmic Web}
\editors{R. van de Weygaert, S. Shandarin, E. Saar \& J. Einasto, eds.}
\begin{document}

\maketitle

\begin{abstract}
Understanding the internal structure and spatial distribution of cosmic voids is crucial when considering them as probes of cosmology. We present recent advances in modeling void density- and velocity-profiles in real space, as well as void two-point statistics in redshift space, by examining voids identified via the watershed transform in state-of-the-art $\Lambda$CDM n-body simulations and mock galaxy catalogs. The simple and universal characteristics that emerge from these statistics indicate the self-similarity of large-scale structure and suggest cosmic voids to be among the most pristine objects to consider for future studies on the nature of dark energy, dark matter and modified gravity.
\keywords{cosmology: large-scale structure of universe, dark matter, cosmological parameters, gravitation, methods: n-body simulations}
\end{abstract}

\firstsection 
\section{Introduction}
A fundamental quantity to describe the structure of cosmic voids in a statistical sense is their spherically averaged density profile. In contrast to the well-known formulas parametrizing density profiles of simulated dark matter halos, rather few models for void density profiles have been developed, mainly focusing on their central regions~(e.g., \cite{Colberg2005,Padilla2005,Ricciardelli2014}), rarely taking into account the compensation walls around voids~(\cite{Paz2013}). We present a simple formula able to accurately describe the real-space density and velocity profile of voids and subvoids of any size and redshift out to large distances from their center~(\cite{Hamaus2014b}).

We further propose a new statistic to be adequate for cosmological applications: void auto-correlations. In contrast to void-galaxy cross- (\cite{Padilla2005,Paz2013}) and galaxy auto-correlations, this statistic is not directly affected by redshift-space distortions, since the galaxy density does not enter in it. Peculiar motions of galaxies only affect this estimator indirectly when voids are identified in redshift space. However, the difference in void positions from real space is expected to be small, as many galaxies are needed to define a single void, which diminishes the net displacement of its center. This property makes the void auto-correlation a very pristine and model-independent statistic to conduct an Alcock-Paczynski (AP) test~(\cite{Hamaus2014c}).

In this study we apply the publicly available Void IDentification and Examination toolkit (VIDE, \cite{Sutter2014b}), which is based on the watershed algorithm ZOBOV (\cite{Neyrinck2008}), to n-body simulations (\cite{Warren2013}) and derived mock galaxy catalogs. The code accounts for the self-similar nature of the cosmic web by arranging a nested hierarchy of voids and subvoids into a tree-like structure.

\section{Results}
\begin{figure*}[!t]
\centering
\resizebox{\hsize}{!}{
\includegraphics[trim=0 0 0 0,clip]{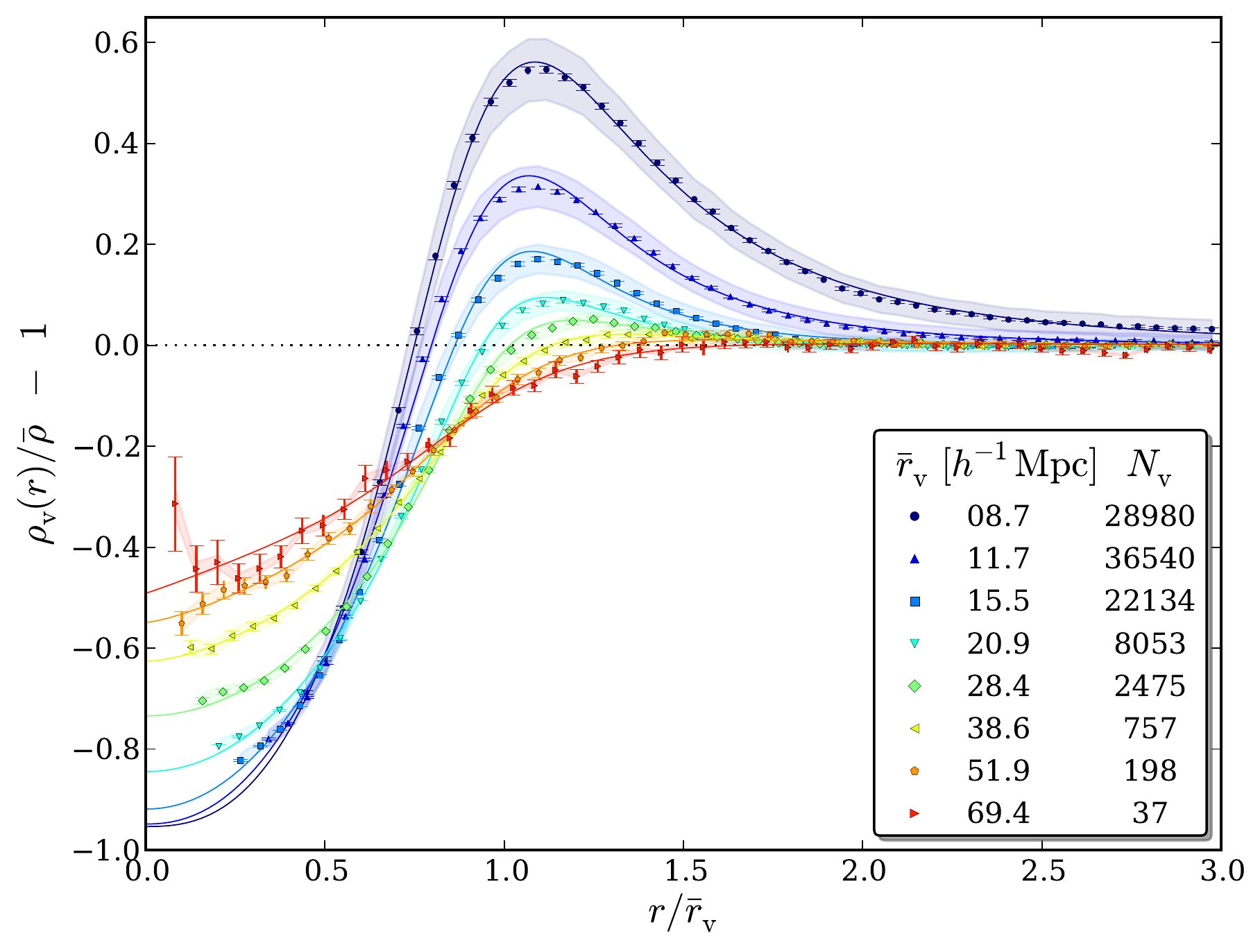}
\includegraphics[trim=0 0 0 0,clip]{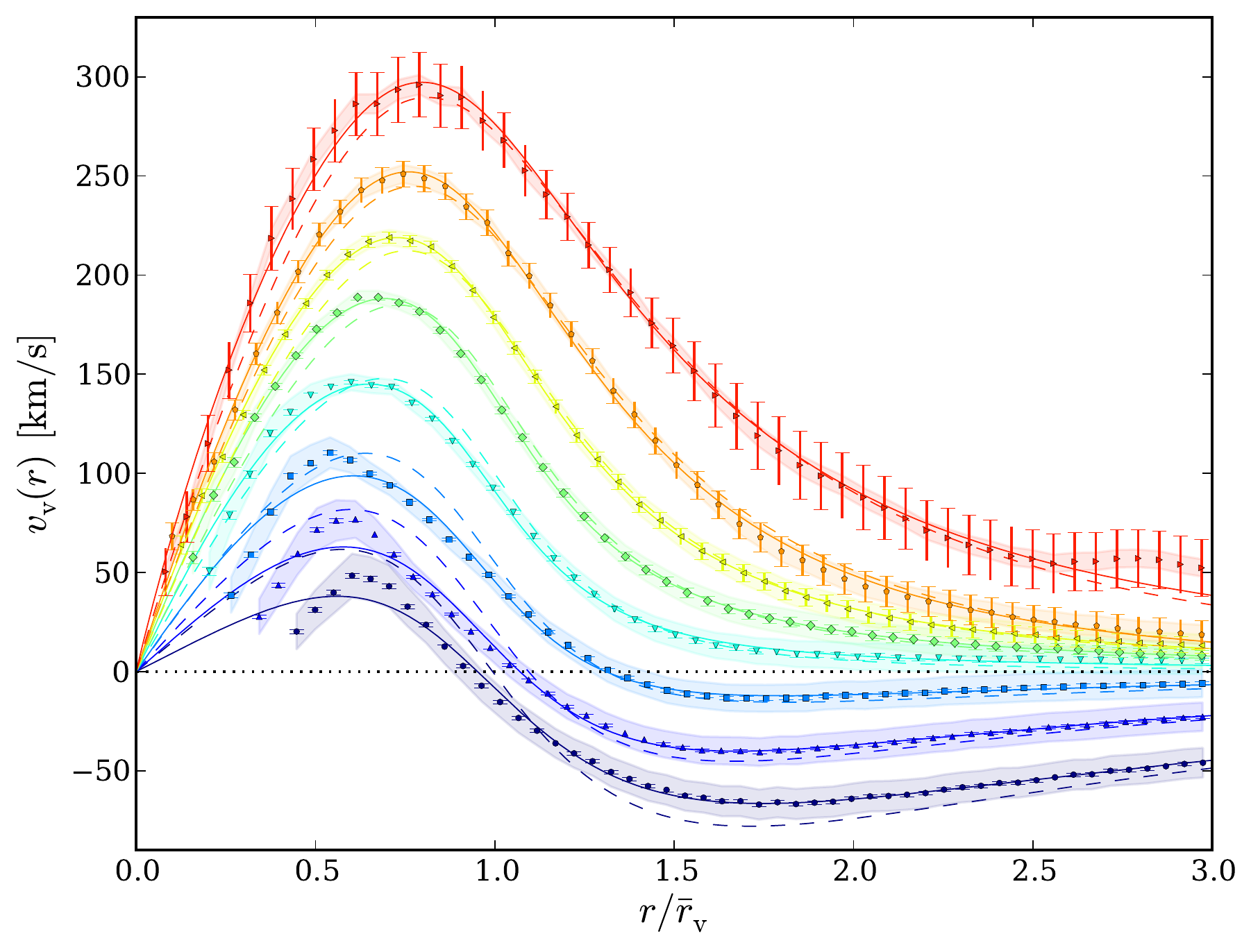}}
\caption{Stacked real-space density (left) and velocity (right) profiles of voids at $z=0$ with mean effective radii and void counts indicated in the inset. Shaded regions depict the standard deviation $\sigma$ within each of the stacks (scaled down by 20 for visibility), while error bars show standard errors on the mean profile $\sigma/\sqrt{N_\void}$. Solid lines represent individual best-fit solutions and dashed lines show linear theory predictions for velocity obtained from the density stacks.}
\label{fig1}
\end{figure*}

{\underline{\it Density structure}}. We define the void density profile as the spherically averaged relative deviation of mass density around a void center from the mean value $\bar{\rho}$ across the Universe, $\rho_\void(r)/\bar{\rho} - 1$. Note that we consider the full void hierarchy including multiple levels of subvoids here, while \cite{Nadathur2014} restrict their analysis to only top-level voids in moderate sampling densities, and apply further restrictive cuts. The results from shell-averaging particles around voids in bins (stacks) of effective radius are shown in the left-hand panel of Fig.~\ref{fig1}. We have conducted a resolution study in order to ensure our density estimation to be unbiased for all the data points shown.

As expected, stacked voids are deeply underdense inside, with their central density increasing with void size. In addition, the variance of underdense regions is suppressed compared to overdense ones, yielding the smallest error bars in the centers of the emptiest voids. However, note that the void-to-void scatter in the profile decreases towards the largest voids, as can be seen from the shaded regions in Fig.~\ref{fig1}. The profiles all exhibit overdense compensation walls with a maximum located slightly outside their mean effective void radius $\bar{r}_\void$, shifting outwards for larger voids. The height of the compensation wall decreases with void size, causing the inner profile slope to become shallower. This trend divides all voids into being either over- or undercompensated, depending on whether the total mass within their compensation wall exceeds or falls behind their missing mass in the center, respectively. We propose a simple empirical formula that accurately captures the properties described above:
\begin{equation}
 \frac{\rho_\void(r)}{\bar{\rho}} - 1 = \dc\,\frac{1-(r/\rs)^\alpha}{1+(r/r_\void)^\beta}\;, \label{dprofile}
\end{equation}
where $\dc$ is the central density contrast, $\rs$ a scale radius at which $\rho_\void=\bar{\rho}$, and $\alpha$ and $\beta$ determine the inner and outer slope of the void's compensation wall, respectively. The best fits of this model to the data are shown as solid lines in the left-hand panel of Fig.~\ref{fig1}.

{\underline{\it Velocity structure}}. The right-hand panel of Fig.~\ref{fig1} depicts the resulting velocity stacks using the same void radius bins. Note that a positive velocity implies outflow of tracer particles from the void center, while a negative one denotes infall. As the largest voids are undercompensated (void-in-void process in \cite{Sheth2004}), they are characterized by outflow in the entire distance range. Small voids exhibit infall velocities, as they are overcompensated (void-in-cloud process). This causes a sign change in their velocity profile around the void's effective radius beyond which matter is flowing onto its compensation wall, ultimately leading to a collapse of the void.

The distinction between over- and undercompensation can directly be inferred from velocities, since only overcompensated voids feature a sign change in their velocity profile, while undercompensated ones do not. Consequently, the flow of tracer particles around precisely compensated voids vanishes already at a finite distance to the void center and remains zero outwards. We denote the effective radius of such voids the \emph{compensation scale}. It can also be inferred via clustering analysis in Fourier space, as compensated structures do not generate any large-scale power (\cite{Hamaus2014a}).

In linear theory the velocity profile can be related to the density using $v_\void(r) = -\frac{1}{3}\Omega_\mathrm{m}^\gamma H r\Delta(r)$, where $\Omega_\mathrm{m}$ is the relative matter content in the Universe, $\gamma\simeq0.55$ the growth index of matter perturbations, $H$ the Hubble constant, and $\Delta(r)$ the integrated density contrast defined as $\Delta(r) = \frac{3}{r^3}\int_0^r\left[\rho_\void(q)/\bar{\rho}-1\right]q^2 \mathrm{d}q$.
With Eq.~(\ref{dprofile}), this yields
\begin{equation}
 \Delta(r) = \dc\,{}_2F_1\!\!\left[1,\frac{3}{\beta},\frac{3}{\beta}+1,-(r/r_\void)^\beta\right]-\frac{3\dc(r/\rs)^\alpha}{\alpha+3}\,{}_2F_1\!\!\left[1,\frac{\alpha+3}{\beta},\frac{\alpha+3}{\beta}+1,-(r/r_\void)^\beta\right], \label{intprofile}
\end{equation}
where ${}_2F_1$ is the Gauss hypergeometric function. We can use this analytic formula to fit the velocity profiles obtained from our simulations; the results are shown as solid lines in the right-hand panel of Fig.~\ref{fig1}. As for the density profiles, the quality of the fits is remarkable, especially for large voids. Only the interiors of smaller voids show stronger discrepancy, which is mainly due to the decreasing validity of the linear theory relation between density and velocity. We obtain best-fit parameter values that are very similar to the ones resulting from the density stacks above. In fact, evaluating the velocity profile at the best-fit parameters obtained from the density stacks yields almost identical results, as indicated by the dashed lines in Fig.~\ref{fig1}.

{\underline{\it Geometry}}. Although individual voids exhibit arbitrary shapes and internal structures, their ensemble average must obey statistical isotropy, if the cosmological principle holds. In observations, void density profiles are constructed by alignment of the barycenters of each individual void identified in a galaxy survey, and by histogramming the distribution of galaxies around this center. This is equivalent to the void-galaxy cross-correlation function and hence a two-point statistic like the galaxy correlation function, which can be exploited for the inference of cosmological parameters. However, these stacked voids are subject to redshift-space distortions, which are inherent to the galaxies used to construct their profiles~(\cite{Padilla2005,Sutter2012,Paz2013,Hamaus2014c}). 

\begin{figure*}[!t]
\centering
\resizebox{\hsize}{!}{
\includegraphics[trim=0 0 0 0,clip]{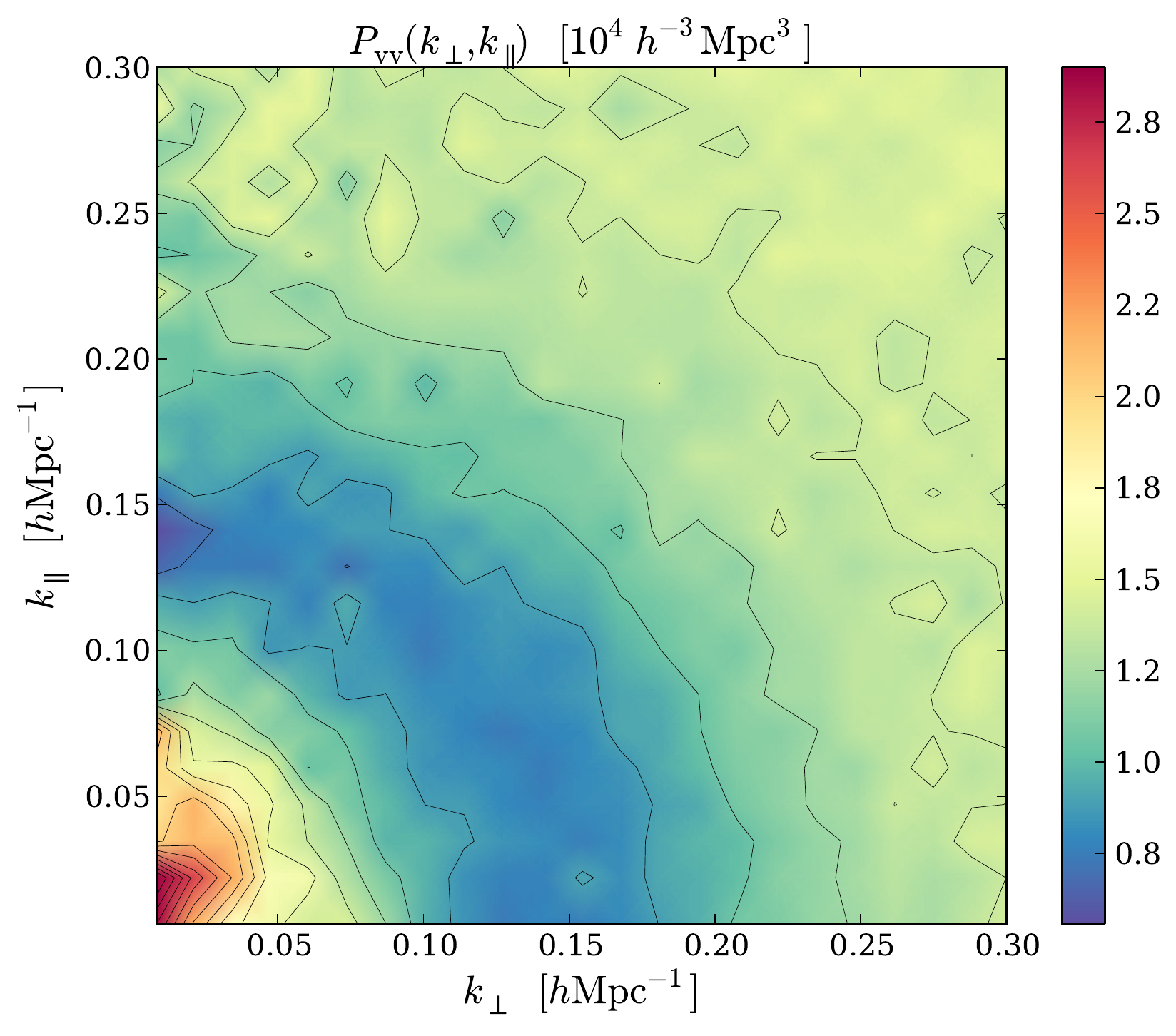}
\includegraphics[trim=0 0 0 0,clip]{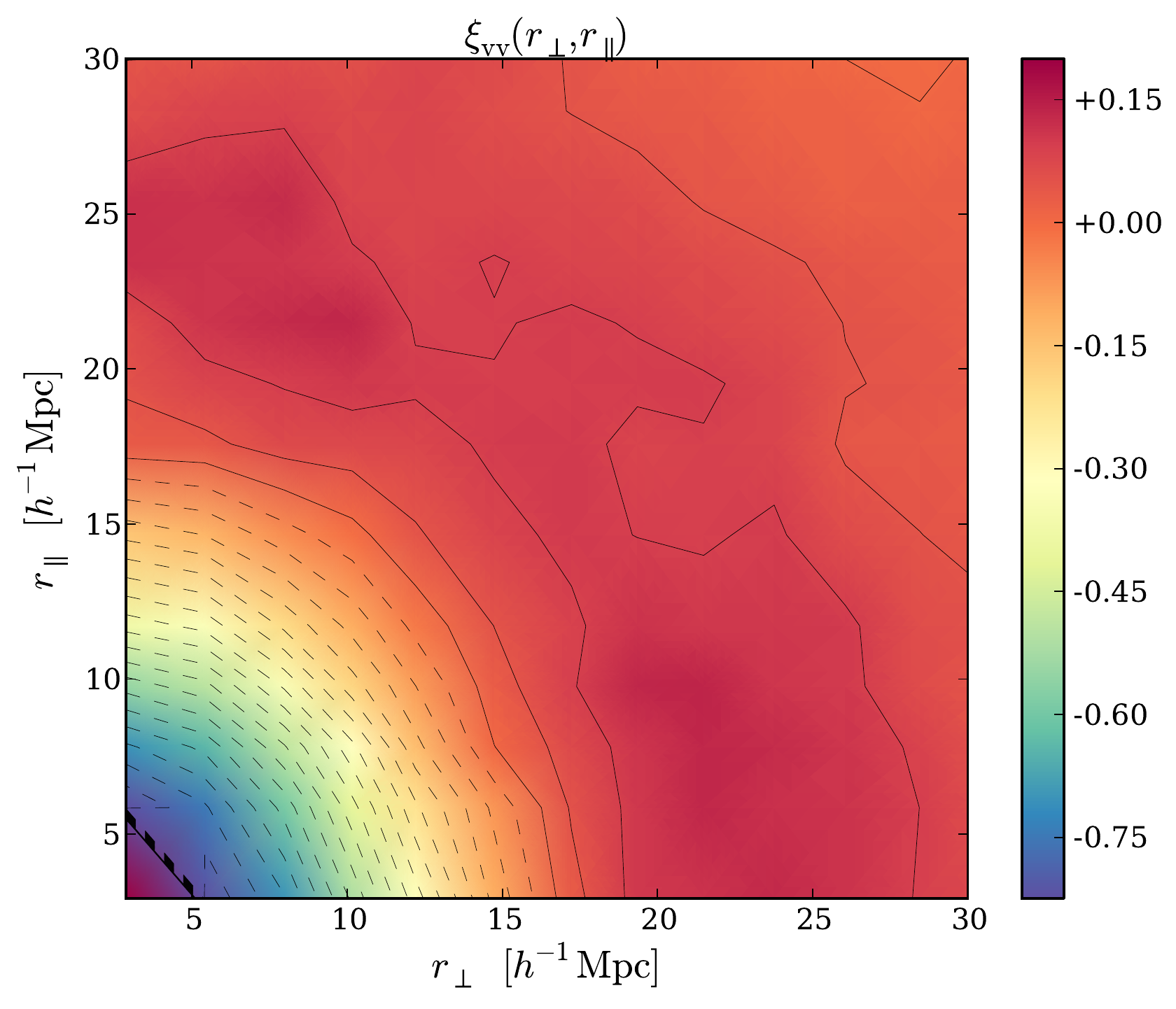}}
\caption{Two-dimensional void power spectrum (left) and void correlation function (right) from galaxies in redshift space at $z=0$. Solid lines show positive, dashed lines negative contours.}
\label{fig2}
\end{figure*}

The two-point statistics (auto-power spectrum and auto-correlation function) for voids found in mock galaxies in redshift space are shown in Fig.~\ref{fig2}. Dynamic distortions evidently play a minor role here, as the contour lines appear to be fairly circular. Residual anisotropies are consistent with being random fluctuations due to the lower number count of voids compared to galaxies in a given survey volume, but no systematic trends are manifest. The compensation wall of the void auto-correlation function extends to twice the effective void radius $\bar{r}_\void$, because of mutual void exclusion. 
This feature manifests itself as a ring of suppressed power in Fourier space at a scale $k\simeq\pi/2\bar{r}_\void$ (\cite{Hamaus2014c}). Existing studies so far have only considered galaxy auto- and void-galaxy cross-correlations for applications of the AP test (\cite{Sutter2012}), but void auto-correlations can be readily obtained from existing data as well and provide additional information on the inferred geometry of large-scale structure. Moreover, systematic uncertainties and biases from redshift-space distortions appear to be negligible in void auto-correlations, which potentially makes this statistic the cleanest one for detecting geometric distortions.

\section{Conclusions}
There are a number of cosmological applications to make use of the presented functional form of the average void density profile, for example, studies of gravitational (anti)lensing that directly probe the projected mass distribution around voids, which in turn may serve as a tool for constraining models of dark matter, dark energy, and modified gravity. 
This is thanks to the universal nature of Eq.~(\ref{dprofile}), which even describes voids in the distribution of galaxies remarkably well, as demonstrated in \cite{Sutter2014a}. That paper further pointed out that the impact of tracer sparsity and bias on the definition of voids can be accounted for by simple rescalings of voids.

Moreover, void auto-correlations are a promising statistic for applications of the AP test and constraining the expansion history of the Universe. In addition to galaxy auto- and void-galaxy cross-correlations, they provide complementary information on cosmological parameters, while at the same time being least affected by systematic effects from redshift-space distortions. Our findings corroborate other indications that cosmic voids may indeed offer new and complementary approaches to modeling fundamental aspects of the large-scale structure of our Universe.

\end{document}